\documentclass[aps,prl,groupedaddress,showpacs,showkeys,floatfix,dvipdfmx]{revtex4-1}
\usepackage{wrapfig}
\usepackage{graphicx}
\usepackage{colordvi}
\usepackage{color}
\usepackage{amsmath}
\usepackage{amssymb}
 \usepackage{comment}
\usepackage{natbib}
\usepackage{ulem}
\usepackage[small, bf, justification=raggedright,format=plain]{caption}

\newcommand{\ve}[1]{\mbox{\boldmath $#1$}}

\usepackage{color}

\begin{document}

\title{
Active Deformable Cells Undergo Cell Shape Transition \\Associated with Percolation of Topological Defects
}

\author{Nen Saito}
\email[]{nensaito@hiroshima-u.ac.jp}
\affiliation{
Graduate School of Integrated Sciences for Life, Hiroshima University, 1-3-1 Kagamiyama, Higashi-Hiroshima City, Hiroshima, 739-8528, Japan;
Exploratory Research Center on Life and Living Systems, National Institutes of Natural Sciences, 5-1 Higashiyama, Myodaiji-cho, Okazaki, Aichi
444-8787, Japan;
Universal Biology Institute, The University of Tokyo, 7-3-1 Hongo, Tokyo, 113-0033, Japan. }

\author{Shuji Ishihara}
\affiliation{
Graduate School of Arts and Sciences, The University of Tokyo, Komaba 3-8-1, Meguro-ku, Tokyo 153-8902, Japan;+
Universal Biology Institute, The University of Tokyo, Komaba 3-8-1, Meguro-ku, Tokyo 153-8902, Japan
}

\date{\today}
\begin{abstract}

Cell deformability is an essential determinant for tissue-scale mechanical nature, such as fluidity and rigidity, and is thus crucial for understanding tissue homeostasis and stable developmental processes. However, numerical simulations for the collective dynamics of cells with arbitral cell deformations akin to mesenchymal, ameboid, and epithelial cells in a non-confluent situation need high computational costs and are still challenging. Here we propose a new method that allows us to study significantly larger numbers of cells than existing methods. Using the method, we investigated the densely packed active cell population interacting via excluded volume interactions, and discovered the emergence of two fluid phases in deformable cell populations, a soft-fluid phase with drastically deformed cell shapes and a fluid phase with circular cell shapes. The transition between these two phases is characterized by the percolation of topological defects, which is experimentally testable.

\end{abstract}
\maketitle

\section{Introduction}
A population of interacting self-propelled agents, often referred to as ``active matter'', exhibits collective movement without external cues, as exemplified by migrating cell populations~\cite{marchetti2013hydrodynamics} or flocks of birds~\cite{cavagna2014bird,marchetti2013hydrodynamics}. 
Even with only excluded volume interactions, various phenomena characteristic to non-equilibrium active systems have been discovered, including giant density fluctuations~\cite{fily2012athermal}, motility-induced phase separation~\cite{cates2015motility}, and active jamming~\cite{henkes2011active}. The shapes of the particles can alter the collective order; the motility-induced phase separation observed in spherical particles~\cite{cates2015motility}
vanishes in slightly non-spherical particles~\cite{grossmann2020particle}. 
A self-propelled rod exhibits the local orientational order~\cite{bar2020self}, and more complex collective patterns appear in more complex shapes~\cite{denk2016active,spellings2015shape,wensink2014controlling,bar2020self}.

For the collective motions of cells, individual cells exhibit significant deformability, which was not considered in conventional models of active matter~\cite{fily2012athermal,cates2015motility,henkes2011active,grossmann2020particle,bar2020self,denk2016active,spellings2015shape,wensink2014controlling,bar2020self}. 
This deformability of the cells manifests in a densely packed situation and can drastically alter the tissue scale rheological properties, such as elasticity or viscosity, by deforming the cells themselves or
facilitating cell rearrangements, and is closely linked to biological functions in embryogenesis, wound healing, and cancer invasion. The population of deformable active particles needs to be addressed to understand the underlying principle behind tissue organization and homeostasis.

Epithelial cells forming a confluent monolayer are well described by the cell-vertex model, in which the cells are densely packed with no gaps, and their shapes are approximated by polygons.
The vertex model was recently used to explain the experimentally observed rigidity transition~\cite{angelini2011glass} from the 
a glassy phase, where the movements of cells are nearly frozen
, to a fluidic phase, where frequent cell rearrangements result in tissue-scale collective motion~\cite{bi2015density}.
Despite the constant density with a volume fraction of one, the vertex model undergoes a rigidity transition, where the deformability of cells triggers a cascade of cell rearrangements. 
The transition occurs when the target cell shape index $q$ defined by $\mbox{(cell perimeter)}/\sqrt{\mbox{(cell area)}}$ exceeds $q=3.81$. 
The self-propelled Voronoi model (SPV) that is almost equivalent to the vertex model, including the self-propelled force~\cite{bi2016motility} also showed the same rigidity transition at $\langle q \rangle =3.81$, where $\langle q \rangle $ is the average value of the shape index for each cell.
These studies suggest that the rigidity transition is conditioned by the geometric parameter $q$: 
which is also supported experimentally~\cite{park2015unjamming}.

On the other hand, it is yet to be well understood how cells other than epithelial cells, such as mesenchymal cells, for whom polygonal approximation is inappropriate, behave at high density.
Although the rigidity transition in epithelial cells was initially studied in the context of epithelial-mesenchymal transition (EMT), experimental evidence suggests that EMT does not correspond to rigidity transition~\cite{mitchel2020primary}. 
If so, it is important to clarify how the mesenchymal cell population acquires fluidity and mobility that often appear in biological processes
and how they differ from the transition in epithelial cells~\cite{bi2015density,bi2016motility}.
To simulate cells with non-polygonal shapes, the phase-field model with self-propulsion was used to explore cells at high density~\cite{loewe2020solid}, and it was shown that
a rigidity transition similar to the vertex model occurs merely by considering the excluded volume interaction. A model of foam-like deformable cells with cell adhesion was also investigated~\cite{kim2021embryonic}, where strong cell adhesion mediates the rigidity transition. For both studies, 
the rigidity transition occurred at a similar $\langle q \rangle$ value in the vertex model. 
However, these models can only handle a smaller number of cells than the vertex model, and it remains to be determined whether a mesenchyme-specific phase exists.

Here, we propose the Fourier contour cell model that allows us to handle up to $10^4$ deformable cells on a single CPU. The cells in the model have a round shape when isolated, whereas, in a high-density situation, 
the model can describe drastic cell deformation caused by the balance between the cell surface tension and excluded volume interactions. We demonstrate that a novel fluid phase specific to mesenchyme-like cells appears through excluded volume interactions and self-propulsion of each cell. The observed phase transitions were accompanied by
the percolation of topological defects, providing a new perspective on mesenchymal cell dynamics that is experimentally verifiable.

\section{Model}
Deformable cells interacting in two-dimensional space are considered here. 
We propose the Fourier contour cell model in which the cell contour is expressed by polar coordinates $R(\theta)$ (Fig.~1a). The cell contour $R(\theta)$ for $i$-th cell centered at the origin is expressed by a Fourier expansion up to $M$-th order, as follows:
\begin{eqnarray}
R^{i}(\theta)=R_0 \sum_{n=0}^{M} \left[ a_n ^{i} \cos n(\theta -\theta^{i}) +b_n ^{i} \sin n(\theta-\theta^{i}) \right]~,
\end{eqnarray}
where $\{a_n ^{i}\}$ and $\{b_n ^{i}\}$ are Fourier coefficients, $R_0$ is a constant parameter with radius dimensions, and $\theta^{i}$ denotes the orientation of the $i$-th cell, which corresponds to the self-propulsion direction of the cell.
In this study, we adopt $M=6$.
Considering the constraints that the cell area $\pi R_0^2$ is constant and that the center of the polar coordinates coincides with the cell centroid, 
$a_0^i$, $a_1^i$ and $b_1^i$ are determined as $a_0^i=\sqrt{1-\sum_{n=1}^{M}[(a_n^i)^2+(b_n^i)^2]/2}$ and $a_1^i=b_1^i=0$, respectively. 
Furthermore, we impose a constraint $\sum_{n=1}^{M}\sqrt{(a_n^i)^2+(b_n^i)^2}<\sqrt{2/3}$ to avoid self-crossing of the contour (see Supplemental Materials). 
The variables for describing $i$-th cell are composed of  the centroid position $\ve{r}_c^{i}$, orientation of the self-propulsion $\theta^{i}$, 
and Fourier coefficients to describe shapes $\{a_n^{i}\}$ and $\{b_n ^{i}\}$~($n=2,3,\ldots,M$). 

As the simplest interaction, only the excluded volume effect is considered.
The interaction Hamiltonian $\mathcal{H}_{int}$ between $i$-th and $j$-th cells is given as an energetic penalty against the overlapped area. 
To compute this overlap, the field representation of the cell, originally used to represent the elliptical shape of the cell~\cite{grossmann2020particle}, was applied to describe deformable cells.
The interior region of $i$th cell is described by $\phi^i (\ve{r})\sim 1$ (Fig.~1b) using $\phi^i (\ve{r})$ calculated as follows:
\begin{eqnarray}
	\phi^i(\ve{r})=\frac{1}{2} + \frac{1}{2}\tanh \left(\frac{R^i\left(\theta(\ve{r},\ve{r}_c^{i}) \right)- \Delta^i(\ve{r},\ve{r}_c^{i})}{\epsilon/2}\right) ~~~
\end{eqnarray}
where $\theta(\ve{r},\ve{r}_c^{i})$ is the angle between $\ve{r}-\ve{r}_c^{i}$ and $x$-axis, and 
$\Delta^i(\ve{r},\ve{r}_c^{i})=|\ve{r}-\ve{r}_c^i |$. 
With small $\epsilon$, 
function $\phi^i (\ve{r})$ sharply rises to $1$ inside the cell (i.e., $\Delta^i<R^i$) and drops to $0$ outside the cell (Fig.~1b).
Interface width between $\phi^i=0$ and $\phi^i=1$ domains
is controlled by $\epsilon$, which is considered infinitesimal below.
The interaction Hamiltonian is then given by $\mathcal{H}_{int}=\sum_{i<j} \int d\ve{r} \phi^i\phi^j$ where we set the coefficient to unity.
We also incorporate the energy for penalizing interface length associated with the membrane tension and obtain the total Hamiltonian $\mathcal{H}=\mathcal{H}_{int}+\eta \int_{-\pi}^{\pi} \sqrt{(R^i)^2+(R^i{}^\prime )^2}d\theta$, where $\eta$ is the tension parameter, and $R^i {}^\prime$ denotes the derivative of $R^i$ with respect to $\theta$. 
The second term does not depend on $\ve{r}_c$ and $\theta^i$.
\begin{figure}[t]
\includegraphics[width=9.5cm,pagebox=cropbox,clip]{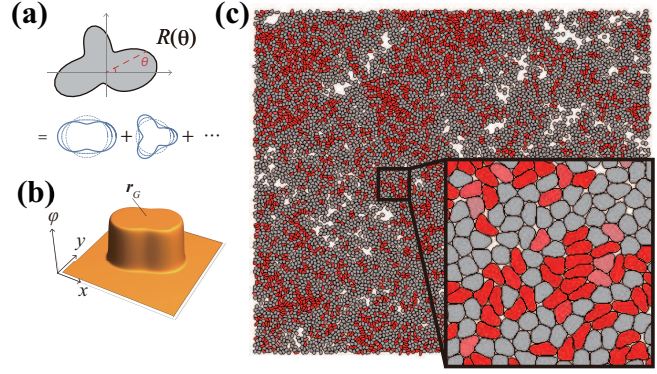}
\caption{
The schematic illustrations of the proposed model. (a) The polar coordinate representation of the cell contour is decomposed by the Fourier expansion into the independent Fourier modes. (b) The field representation of the cell by $\phi$ in which $\phi$ takes $\phi=1$ inside and $\phi=0$ outside of the cell. (c) A snapshot of the simulation with $10^4$ particles with volume density $0.8$. Total simulation time steps are $5\times 10^6$. The red color represents the shape-index (perimeter/$\sqrt{\mbox{area}})$ for each particle. 
 }\label{fig:fig1}
\end{figure}

\begin{figure*}[ht]
\centering
\includegraphics[width=18cm,pagebox=cropbox,clip]{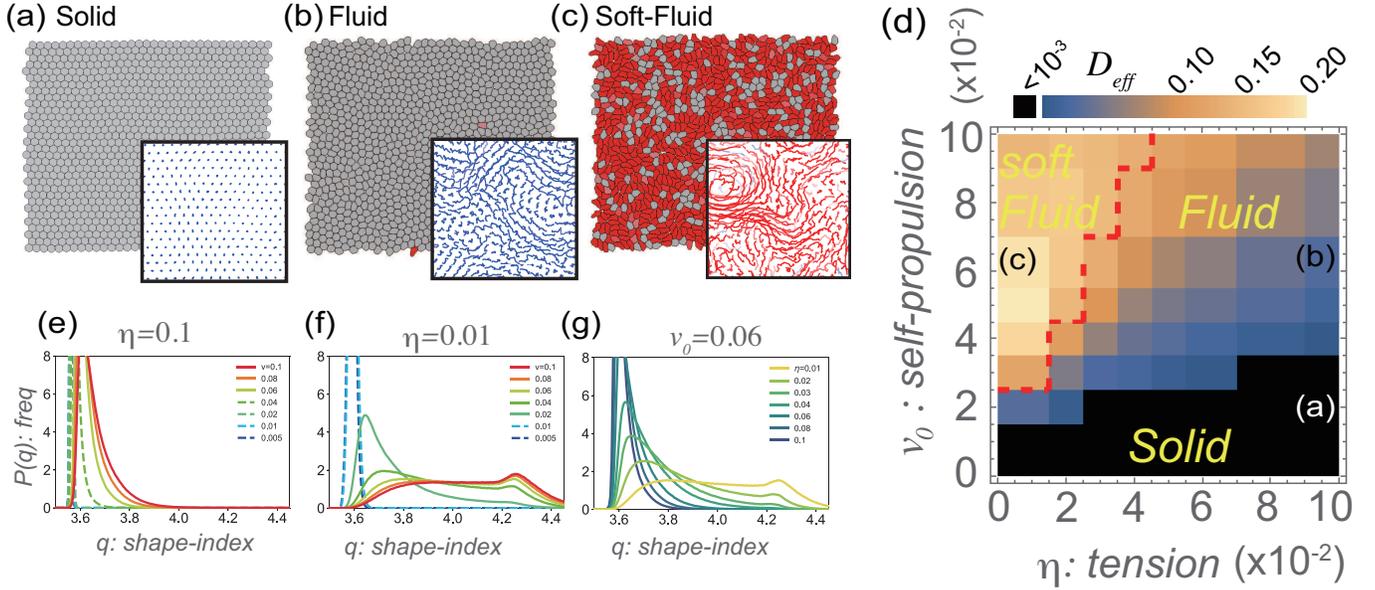}
\caption{Phases for the proposed model. (a-c) The snapshot of the simulation for the solid phase~(a), the fluid phase~(b), and the soft fluid phase~(c).  For cells with the shape-index $q>3.81$, the value of $q$ is indicated by the color scale from gray to red. The insets denote particle trajectories within $10^5$ steps, and the color scale from blue to red represents the average of $q$ during the time steps. (d) Phase diagram of the model against the tension $\eta$ vs. the self-propulsion velocity $v$. The color indicates the effective diffusion constant $D_{eff}$ evaluated from simulation with $N=1024$ and $8\times 10^7$ time steps. (e-g) Distributions of the shape index for $\eta=0.1$ (e) and $\eta=0.01$ (f) with varying $v$, and that for $v=0.06$ with varying $\eta$ (g). Thick lines indicate the parameters for the fluid or soft-fluid phase, and the dashed lines represent the parameters for the solid phase.
} \label{fig:fig1}
\end{figure*}

The time-evolution equation for $\ve{r}^i_c$ is derived from the functional derivative of $\mathcal{H}$ 
with the self-propulsion term $\ve{v}^i=(v_0\cos\theta^i,v_0\sin\theta^i)$, 
which is often introduced for representing active motion of cells~\cite{fily2012athermal,cates2015motility,henkes2011active,grossmann2020particle,bar2020self,denk2016active,spellings2015shape,wensink2014controlling,bar2020self}:
\begin{eqnarray}
	\dot{\ve{r}}_c^{(i)}&=&\ve{v}^i+\frac{\delta\mathcal{H}}{\delta \phi^i} \nabla\phi^i =\ve{v}^i+\sum_{j\neq i}\int d\ve{r} \phi^j |\nabla \phi^i|\frac{\nabla \phi^i}{|\nabla \phi^i|} \label{eq:dotR_1}
\end{eqnarray}
By taking the sharp interface limit $\epsilon \to 0$, $|\nabla\phi^i|$ becomes the surface delta function $\delta_s(\ve{r}-R^i)$ (see supplemental text), and 
the second term is replaced by an integral along the cell contour $\sum_{j\neq i}\oint ds \phi^j \ve{n}^i$
where $\ve{n}^i$ denotes the normal vector of the contour. By using $\ve{e}_\theta=(\cos\theta, \sin\theta)$ and $\ve{e}_{\perp}=(-\sin\theta,\cos\theta)$, 
$\dot{\ve{r}}_c^{(i)}$ is given by
\begin{eqnarray}
	\dot{\ve{r}}_c^{i}&=&\ve{v}^i+\sum_{j\neq i}\!\int_0^{2\pi} \!\!\!\!d\theta \phi^j\! \left[ R^i {}'(\theta)\ve{e}_{\perp}\!\!-\! R^i(\theta)\ve{e}_\theta\right]~~~
\end{eqnarray}
At the last integral, $\phi^j=1$ when the position $\ve{r}_c^i+R^i\ve{e}_\theta$ on the contour in the $\theta$-direction of the $i$-th cell
is occupied by the $j$th cell, and $\phi^j=0$ otherwise.
The time evolutions of $\theta^i$, $\{ a_n\}$ and $\{ b_n\}$ ($n=2,3,\ldots,M$) are obtained in the same manner as $\dot{\theta^i}\propto -\delta \mathcal{H}/\delta \theta+\sqrt{2D_r}\xi^i$, $\dot{a}_n\propto -\delta \mathcal{H}/\delta a_n$ and $\dot{b}_n\propto -\delta \mathcal{H}/\delta b_n$. 
Note that the tension term in the Hamiltonian affects only the equations for $\{ a_n\}$ and $\{ b_n\}$.
For simplicity, we incorporate a noise term $\sqrt{2D_r}\xi^i$, where $\xi^i$ represents a normalized white Gaussian noise, only into the equation of $\dot{\theta}^i$.
The numerical integration along the cell contour $\theta$ was performed by discretizing $0\le \theta \le 2\pi$ by 40 points, whereas the integration with respect to time $t$ was performed using the Euler-Maruyama method with $\Delta t=5 \times 10^{-3}$.
The dynamics in the centroid $\ve{r}^i_c$, orientation $\theta^i$ and cell shape $\{ a_n^i, b_n^i\}$ are described by a set of ordinary differential equations (ODEs):
where the two-dimensional numerical integration in Eq.~(\ref{eq:dotR_1}) is reduced to a one-dimensional integration, which significantly reduces the computation time. 
Figure~1(c) shows a representative simulation snapshot of $N=10^4$ active deformable particles (see also VIDEO1 in Supplemental Material), which may not be feasible using the phase-field method~\cite{loewe2020solid,lober2015collisions,nonomura2012study}.

\section{Results}
Using the proposed model, we study active deformable particles densely packed with volume fraction $\rho =0.95$
and interact with each other solely through excluded volume interactions.
The simulation box size was set as $L_x:L_y = 2:\sqrt{3}$ and the particles were initially positioned 
on a perfect hexagonal lattice with random $\theta^i$.  
Although up to $N=10^4$ particles were handled, calculations were performed mainly with $N=1024$ when searching the parameter space.  

Representative snapshots are shown in Figs.~2(a)-(c). At a small self-propulsion $v_0$ and a large tension $\eta$, the system exhibits a solid phase (Fig.~2a) 
where the particles form a hexagonal lattice and hardly show rearrangements (particle trajectories are shown in the inset of Fig.~2a; see also VIDEO2 in Supplemental Material). 
Accordingly, the mean square displacement (MSD) measured from the trajectories of the cell centroids 
exhibits a saturation curve (Fig.~S1; $\eta$ = 0.08, 0.1). The particles in this phase were nearly circular without a large deformation (Fig.~2a). 
At large $v_0$ and large $\eta$, the system is in the fluid phase (Fig.~2b), where the particle positions are frequently rearranged (Fig.~2b inset; see also VIDEO3 in Supplemental Material) 
and the MSD curve exhibited linear growth, indicating diffusive dynamics 
of cells (Fig.~S1; $\eta$ = 0.02 - 0.06). 
The particle shapes remained almost circular (Fig.~2b; the degree of deformation is denoted by red color). 
At a small $\eta$ and large $v_0$, where the particles are easily deformed, 
the third phase, which we call the soft-fluid phase, emerges (Fig.~2c; VIDEO4 in Supplemental Material). 
In this phase, largely deformed and relatively circular particles coexist, and they exhibit fluidic characteristics, such as
frequent rearrangements (Fig.~2c, inset) and a linearly increasing MSD curve (Fig.~S1; $\eta$ = 0.01).

\begin{figure}[!t]
\centering
\includegraphics[width=8cm,pagebox=cropbox,clip]{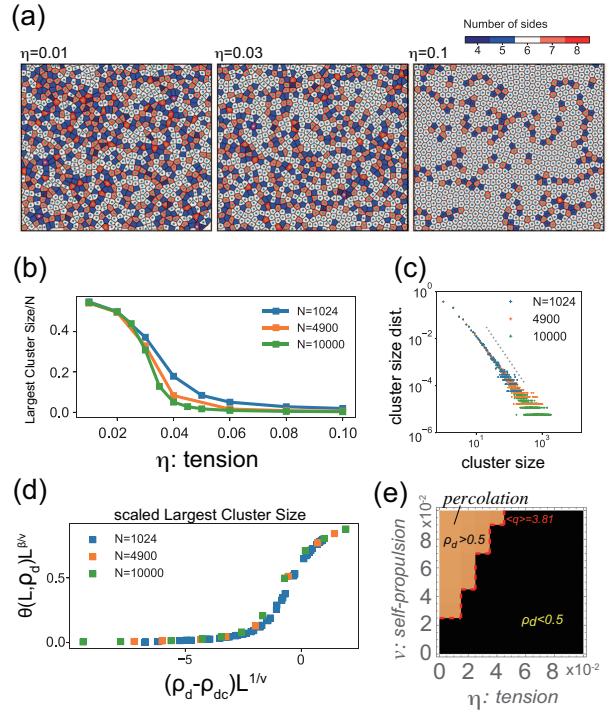}
\caption{
Percolation of the topological defects. (a)~Snapshots of the topological defects for different $\eta$ (top) and the ratio of the number of particles belonging to the largest cluster to the total number of cells $N$ for $v=0.06$ and $N=1024, 4900, 10000$ (bottom). (b) The cluster size distribution for $v=0.06$ and $\eta=0.03$. The dotted line denotes $\propto x^{-187/91}$. (c) The scaled largest cluster size is plotted against the scaled defect ratio $\rho_d$, where the percolation threshold $\rho_d=1/2$, and the scaling exponents $v=4/3$ and $\beta=5/36$ are used. See also Fig.~S7 for the plot without the scaling. 
(d) Parameter region for the defect percolation that is in good agreement with the line for $\langle q \rangle =3.81$.
} \label{fig:defects}
\end{figure}

Figure~2(d) illustrates the phase diagrams for $\eta$ and $v_0$. Color represents the effective diffusion coefficient $D_{\rm eff}=D_{\rm s}/D_0$, where $D_{0} = v_0^2/2D_{\rm r}$ is the diffusion coefficient for an isolated cell.
and $\displaystyle{D_{\rm s}=\lim_{t\to\infty} \left(\langle (\Delta r)^2 \rangle -\langle \Delta r \rangle ^2 \right)/4t}$ were estimated from the trajectories of the cell centroids.
The solid phase is determined by the criterion $D_{\rm eff}<10^{-3}$, which is consistent with the MSD curves showing saturation (Fig.~S1 and S2). 
The solid-fluid phase boundary in Fig.~2(d) indicates that the rigidity-fluidity transition can occur not only by changing the cell propulsion $v_0$ but also by changing the cell deformability $\eta$.
For $v_0=0.03$, $D_{\rm eff}$ varies abruptly at the boundary, and the MSD curves (Fig.~S1) change from a diffusion line ($\eta\le 0.06$) to a saturating curve ($\eta\ge 0.08$).
The phase boundary is also associated with a decrease in the hexatic order parameter $|\Psi_6|$ (Fig.~S3) defined by $\Psi_6(\ve{r})=\langle \sum^{n_i}_{j=1} e^{i6\theta_{ij}}/n_i \rangle$, where $\theta_{ij}$ is the angle of the link connecting $i$-th cell's centroid to the adjacent $j$-th cell's centroid, $n_i$ is the number of adjacent pairs, and $\langle \rangle $ denotes the average over all cells.

The phase boundary between the fluid and soft-fluid is not straightforwardly determined. 
As these two fluid phases have distinct cell shape characteristics, the distribution of the shape index $q$ is measured (Fig.~2e-g) to characterize the phases. 
For a large $\eta$ (high tension), the shape distribution displays a sharp, unique peak at $q<3.81$ 
regardless of whether it was in the solid phase (small $v_0$; Fig.~2e, dashed lines) or fluid phase (large $v_0$; Fig.~2e, thick lines).
On the other hand, for a small $\eta$ (low tension), a sharply peaked distribution at $q<3.81$ in the solid phase (small $v_0$; Fig.~2e, dashed lines) 
turns into a long-tailed distribution as $v_0$ increases and then into a bimodal distribution (Fig.~2e, thick lines).
Changes in the distribution from unimodal to bimodal also occur without crossing the solid phase. 
For a fixed value of $v_0=0.06$, we observe the appearance of a bimodal distribution with a second peak at $q>3.81$ from the unimodal distribution with a peak at $q<3.81$ 
when $\eta$ decreased (Fig.~2g). This bimodal distribution clearly illustrated the coexistence of largely deformed and relatively circular particles. 
Accordingly, the mean $\langle q \rangle$ increases around the region with a small $\eta$ and high $v$ (Fig.~S3).
Based on this change in the distribution and the previously reported transition point in the vertex model~\cite{bi2015density,bi2016motility}, we tentatively defined the phase boundary as the mean shape index $\langle q \rangle =3.81$ (Fig.~2d, dashed red line). This heuristic definition will be justified below.

To further examine the boundary between the fluid and soft-fluid phases, we examined behaviors of topological defects characterized by miscoordinated particles with respect to a perfect hexagonal lattice.
After performing Voronoi tessellation using particle centroids, $\{\ve{r}_c^i \}$,  
defect particles are defined as Voronoi cells with other than six neighboring cells.
Top panels in Fig.~\ref{fig:defects}(a) shows the defect particles indicated by the colors for $v_0 = 0.06$ at $\eta = 0.01, 0.03$, and $0.1$. As $\eta$ decreases from $0.1$ to $0.01$, the number of defect particles increases, and they are connected to form a large cluster (Fig.~\ref{fig:defects}(a)). 
Figure~\ref{fig:defects}(b) demonstrates a clear hallmark of the percolation transition
the largest cluster size of the connected defect particles exhibited a sharp transition between $\eta=0.02$ and $0.06$, 
and the transition becomes sharper with increasing system size $N$ with a fixed density. 
Consistently, the size distribution of the connected clusters of the defect particles exhibited a power law near the critical point $\eta=0.03$ for $v_0=0.06$ (Fig.~3c). 
The power exponent is close to the critical exponent for 2D percolation, $187/91$~\cite{stauffer2018introduction}. 
The cluster size distributions for the other values of $\eta$ are shown in Fig.~S5.
In contrast to the largest cluster size, the defect ratio $\rho_d$, defined as the average number of defect particles/$N^2$, 
does not exhibit a significant change, such as a transition. 
During a decrease in $\eta$,
$\rho_d$ increases smoothly and its curve does not change with increasing $N$ (Fig.~S6). 
The scatterplot of the defect ratio vs. the size of the largest cluster for all examined parameters collapses to a master curve (Fig.~3d; see also Fig.~S7 for the unscaled version) by applying the percolation threshold $\rho_{d\rm c} = 0.5$ for two-dimensional site percolation in the triangular lattice 
and its scaling exponents $v=4/3$ and $\beta=5/36$.
This result clearly demonstrates the existence of percolation transition at $\rho_{d\rm c}$ = 0.5.
Interestingly, the parameter region with $\rho_d > 0.5$ in the phase diagram coincided with the region with $\langle q \rangle >3.81$ (Fig.~3e; see also Fig.~S4), indicating that the percolation transition separates the soft-fluid phase from the fluid phase.

One possible interpretation is that the fluid/soft-fluid transition found in the proposed model corresponds to the hexatic/fluid phase transition in Kosterlitz-Thouless-
Halperin-Nelson-Young (KTHNY) theory. 
According to the KTHNY theory~\cite{strandburg1988two,bernard2011two}, two-dimensional crystals display two-step melting. 
At the first transition, the quasi-long-range order of a crystal is destroyed by the dissociation of the thermally excited pairs of dislocations, which is a linear crystallographic defect, while the six-fold orientational order of the lattice is maintained. This phase is referred to as the hexatic phase. 
Subsequently, the hexatic phase undergoes the second phase transition to the isotropic fluid phase by losing the orientational order mediated by the dissociation of disclinations~\cite{strandburg1988two,bernard2011two}, a defect in which the rotational symmetry is broken. This two-step melting was confirmed in passive~\cite{bernard2011two} and active discs~\cite{klamser2018thermodynamic}, as well as in the Voronoi model~\cite{li2018role,pasupalak2020hexatic}.
The question that arises here is whether the fluid and soft-fluid phases in the proposed model correspond to either hexatic or isotropic fluid phases in the KTHNY theory.
We addressed this issue by closely examining the transition from the solid to the fluid phase.
We measured the pair translational correlation function $G_r(\Delta x,0)$~\cite{bernard2011two}, which is calculated using the one-dimensional cut of the two-dimensional histogram $G_r(\Delta \ve{r})$ of the pair distance $\Delta \ve{r}=\ve{r}_j-\ve{r}_i$ of an $i$-$j$ pair of particle centroids, 
and orientational correlation function $G_6(|\Delta \ve{r}|)$ defined by $G_6(|\Delta \ve{r}|)=\langle \Psi_6(\ve{r}_i) \cdot  \Psi_6^* (\ve{r}_j)\rangle$.

\begin{figure*}[th!]
\includegraphics[width=16cm,pagebox=cropbox,clip]{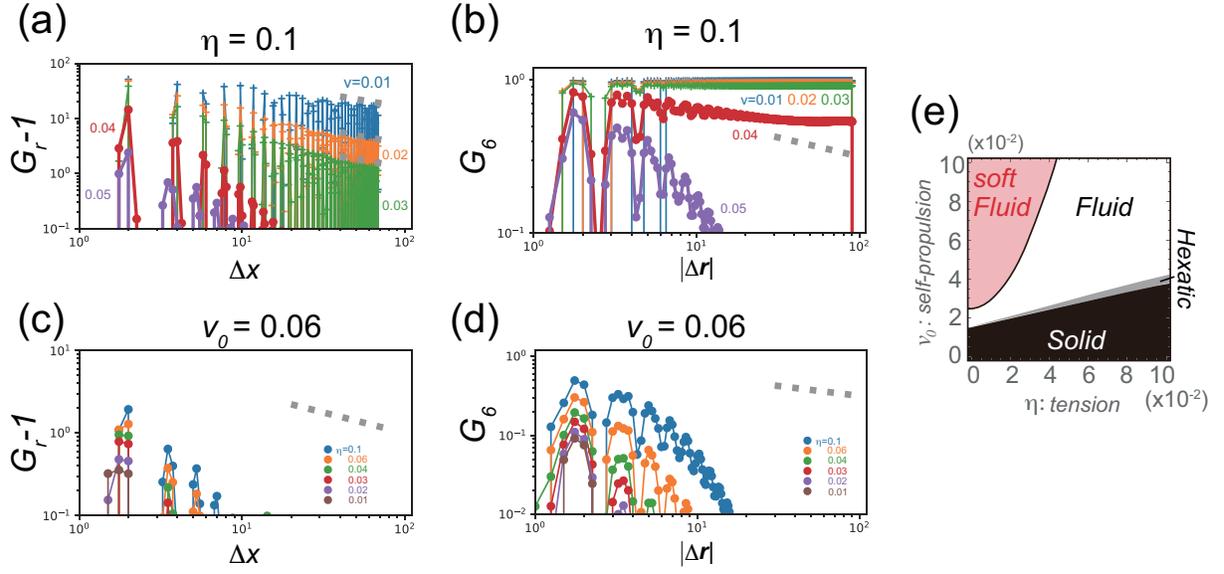}
\caption{The translational and orientational correlation function. 
(a) The translational correlation functions $G_r(\Delta x,0) -1$ for $\eta=0.1$ are plotted from $v=0.01$ (solid phase) to $v=0.05$ (fluid phase). Dashed gray lines represent $G_r-1 \propto (\Delta x)^{-1/3}$. (b) The orientational correlation functions $G_6(|\Delta \ve{r}|)$ for $\eta=0.1$ are plotted from $v=0.01$ (solid phase) to $v=0.05$ (fluid phase). Dashed gray lines represent $G_6 \propto |\Delta \ve{r}|^{-1/4}$. (c, d) $G_r(\Delta x,0) -1$ (c) and $G_6(|\Delta \ve{r}|)$ (d) for $v=0.06$ are plotted from $\eta=0.1$ (fluid phase) to $\eta=0.01$ (soft-fluid phase). Dashed gray lines represent $G_r-1 \propto (\Delta x)^{-1/3}$ (c) and $G_6 \propto |\Delta \ve{r}|^{-1/4}$ (d). All calculations in (a-d) are based on simulation with $N=4900$ and $2\times10^7$ steps.
(e) Summarized phase diagram.
}
\end{figure*}
We calculate $G_r(\Delta x,0)$ and $G_6(|\Delta \ve{r}|)$ by fixing $\eta = 0.1$ and increasing $v_0$ from $0.01$ to $0.05$. 
As shown in Fig.~4(a), from $v_0=0.01$ to $0.03$, the translational order $G_r(\Delta x,0)$ decays algebraically with exponent $-1/3$, 
corresponding to the solid phase in KTHNY theory. $G_r(\Delta x,0)$ decreases exponentially for $v_0$ from $0.04$ to $0.05$. 
The orientational order $G_6(|\Delta \ve{r}|)$ shown in Fig.~4(b) decays more slowly than $G_6(|\Delta \ve{r}|) \propto |\Delta \ve{r}| ^{-1/4}$ for $v_0$ from $0.01$ to $0.04$, corresponding to solid or hexatic phases, and decreases faster than $ |\Delta \ve{r}| ^{-1/4}$ for $v=0.04$ and $0.05$. 
These results indicate that the solid and fluid phases in the proposed model correspond to the solid and isotropic fluid phases in KTHNY theory, and the hexatic phase exists between them for $v_0=0.04$ and $\eta = 0.1$ (Figs.~4a and b).
During the transition from the fluid phase to the soft-fluid phase, $G_r(\Delta x,0)$ and $G_6(|\Delta \ve{r}|)$ decayed exponentially  
(Figs.~4 (c) and (d); $v_0=0.06$ and $\eta$ is changed), indicating that both translational and orientational orders are already broken. 
From these results, we conclude that both the fluid and soft-fluid phases in the proposed model are classified into the isotropic fluid phase in terms of KTHNY theory, and the soft-fluid phase is a novel phase that can be described by the percolation of 
topological defects but not by breaking orientational or translational orders. This scenario is summarized in the phase diagram shown in Fig.~4(e).

\section{Discussion}
The present study proposes a numerical model of deformable cells based on the Fourier expansion of the cell contour, which
provides a computationally efficient method for describing irregular cell shape deformations similar to mesenchymal and ameboid cells. 
Up to $10^4$ cells were handled using a single CPU, which may not be feasible in the phase-field method~\cite{loewe2020solid,lober2015collisions,nonomura2012study} 
or spring-beads models~\cite{kim2021embryonic}. Similar models based on the expansion of cell contours have also been proposed~\cite{ohta2017dynamics,yamanaka2014formation,menzel2012soft,itino2011collective}; however, 
these models had difficulty simulating a tightly packed situation and torque effects, and thus phenomena such as glassy dynamics and rigidity transition 
do not occur~\cite{itino2011collective}.
Our approach is based on the field representation of cell shape $\phi^i$, which enables simple and efficient computation of the excluded volume effect at high density and allows us to incorporate the torque effect appropriately, which can alter collective behaviors~\cite{hiraiwa2022collision,grossmann2020particle}.

The proposed model demonstrated that density-independent rigidity transition occurs by changing deformability $\eta$ solely from the excluded volume interaction and self-propulsion. Although such a transition was reported in a previous study~\cite{loewe2020solid}, 
we further elucidated the emergence of two types of fluid phases: the soft-fluid phase, where the cell shapes drastically deform, 
and the fluid phase, where individual cells remain almost circular.
The soft-fluid phase is characterized by $\langle q \rangle >3.81$ and is thus similar to the phase that appears in the cell-vertex model~\cite{bi2015density,bi2016motility}, 
whereas the fluid phase seems to correspond to the fluidic phase in the active Brownian particle model (ABP) since the cell shapes remain circular.
These two phases stem from the nature of the proposed model, in which the model displays ABP behavior in the high tension limit, while deformability plays an essential role in low tension.
Interestingly, the solid-fluid transition point $\langle q \rangle =3.81$ in the cell-vertex model corresponds to the liquid-liquid transition point in the proposed model. This difference would be explained by the fact that a fluid phase with $\langle q \rangle <3.81$ cannot appear in the cell-vertex model.

The fluid/soft-fluid transition is well-characterized by the percolation of topological defects. 
Recently, such defects percolation has been studied
in equilibrium disks and ABP systems by changing the packing fraction, and it was found that 
percolation occurs around the hexatic-fluid phase-transition boundary
~\cite{digregorio2022unified}.
This difference between the previous study and our study may be due to the 
density-dependent/independent nature of the transition
or the absence/presence of deformability of cells.
Our finding of percolation in the liquid phase also differs from previously reported percolations related to the rigidity transition, such as density-dependent percolation in embryo genesis~\cite{petridou2021rigidity} or stress field percolation in the epithelial monolayer~\cite{monfared2022stress}.
In our case, the analysis of the translational and orientational order parameters revealed that the fluid/soft-fluid transition does not correspond to 
the hexatic/isotropic fluid transition in KTHNY theory,
and both phases have the characteristics of an isotropic fluid.
Moreover, the hexatic phase exists around the solid/fluid transition point
(see Figs.~4a and b; $\eta=0.04$ and Fig.~4e), indicating that three fluid phases can potentially appear in the active deformable cell population.

The fluidic collective motion of actual cells can be characterized using the present theoretical framework. The packing of cell populations with non-polygonal shapes appears in embryo genesis~\cite{kim2021embryonic} and in vitro experiments on cultured cells with suppressed cell-cell adhesion~\cite{balasubramaniam2021investigating}, and they potentially display fluidic collective motion by cell-softening. The percolation of topological defects can be detected under these experimental conditions and provides insights into the novel fluid/fluid transition in the cell population. 

The proposed model also has the flexibility to incorporate various ingredients, such as cell size differences, cell adhesion, and cell division, which need to be addressed elsewhere. The interaction Hamiltonian is based on the overlapped area between two cells, but its extension to a harder core potential is also possible. Extension to a three-dimensional model using spherical harmonic expansions instead of Fourier series expansions can be considered. Thus, the proposed model, which enables a simple and efficient computation of deformable cells, can be a pivotal tool for studying cell populations.

\section{Acknowledgment}
We would like to acknowledge the helpful discussions with Michio Tateno, Atsushi Ikeda, Hideyuki Mizuno, Norihiro Oyama, Kyohei Takae, and Tetsuya Hiraiwa.
This study was supported in part by the Japan Society for the Promotion of Science (JSPS) KAKENHI(21K03496 and 21H05793 to NS). 
Japan Science and Technology Agency (JST) CREST JPMJCR1923 ( NS and SI). 
and the Cooperative Study Program of Exploratory Research Center on Life and Living Systems (ExCELLS; program No. 20-102, 21-102 to NS)
\bibliographystyle{apsrev4-1}

%

\end{document}